\documentclass[conference]{IEEEtran}
\usepackage[latin1]{inputenc}
\IEEEoverridecommandlockouts
\usepackage{cite}
\usepackage{amsmath,amssymb,amsfonts}
\usepackage{algorithmic}
\usepackage{graphicx}
\usepackage{textcomp}
\usepackage{xcolor}
\usepackage{multirow}
\usepackage{txfonts}

\makeatletter

\begin{document}

\title{Deep Multi-Scale Features Learning for Distorted \\Image Quality Assessment
}

\author{Wei Zhou,~\IEEEmembership{Student Member,~IEEE}, and Zhibo Chen,~\IEEEmembership{Senior Member,~IEEE}
\thanks{W. Zhou and Z. Chen are with the CAS Key Laboratory of Technology in Geo-Spatial Information Processing and Application System, University of Science and Technology of China, Hefei 230027, China, (weichou@mail.ustc.edu.cn; chenzhibo@ustc.edu.cn).}
\thanks{This work was supported in part by NSFC under Grant U1908209, 61632001 and the National Key Research and Development Program of China 2018AAA0101400.}}

\maketitle

\begin{abstract}
Image quality assessment (IQA) aims to estimate human perception based image visual quality. Although existing deep neural networks (DNNs) have shown significant effectiveness for tackling the IQA problem, it still needs to improve the DNN-based quality assessment models by exploiting efficient multi-scale features. In this paper, motivated by the human visual system (HVS) combining multi-scale features for perception, we propose to use pyramid features learning to build a DNN with hierarchical multi-scale features for distorted image quality prediction. Our model is based on both residual maps and distorted images in luminance domain, where the proposed network contains spatial pyramid pooling and feature pyramid from the network structure. Our proposed network is optimized in a deep end-to-end supervision manner. To validate the effectiveness of the proposed method, extensive experiments are conducted on four widely-used image quality assessment databases, demonstrating the superiority of our algorithm.
\end{abstract}


\section{Introduction}
Image quality assessment (IQA) is a fundamental problem in the field of image processing. Original undistorted images are easy to be contaminated during the image processing chain, \emph{i.e.}, acquisition, compression, transmission, reconstruction, and display. It is important to monitor image quality and provide advance notice for end-users. Therefore, automatically estimating perceptual image quality is an urgent need in actual life. The algorithmic IQA predictions are supposed to have a high correlation with subjective ground-truth scores made by a large number of human evaluators (commonly known as the Mean Opinion Score (MOS) \cite{ponomarenko2009tid2008,ponomarenko2013color,zhou20163d,zhou2018visual,zhao2019you}). There have emerged many IQA models for various applications, \emph{e.g.}, image super-resolution \cite{zhou2020blind}, image enhancement \cite{chen2014quality}, image retrieval \cite{sun2018assessing}, and style recognition \cite{lu2015deep}.

Early full-reference (FR) IQA methods, such as peak signal-to-noise ratio (PSNR) calculates image quality according to the intensity difference between reference and distorted images. However, the PSNR ignores visual mechanisms as human observers are the ultimate receivers of input image signals. In order to effectively model the human visual system (HVS) perception, traditional FR IQA methods are developed according to the pre-defined HVS characteristics. Wang \emph{et al.} \cite{wang2004image} propose the structure similarity index (SSIM), in which a similarity index map is constructed by extracting the structural, luminance and contrast information between distorted images and original reference images. Extended variants of SSIM have also been presented, which include the multi-scale SSIM (MS-SSIM) \cite{wang2003multiscale} and the information content weighted SSIM (IW-SSIM) \cite{wang2010information}. Zhang \emph{et al.} \cite{zhang2011fsim} propose a feature similarity index (FSIM) to provide the phase congruency and gradient magnitude feature representation for the HVS perception. Sheikh \emph{et al.} \cite{sheikh2006image} propose the visual information fidelity (VIF), in which the image information fidelity is considered. In \cite{sun2018spsim}, the superpixel-based similarity index (SPSIM) is proposed to evaluate image quality based on the perspective of superpixels.

In recent years, deep learning has shown remarkable ability in various computer vision and image/video processing fields \cite{wang2016cost,jin2019unsupervised,zhao2019infrared}. Except for conventional models, significant progresses have been achieved by exploring deep neural networks (DNNs) for IQA \cite{zhou2019dual,xu2020blind,xu2020binocular}. For example, in \cite{wang2016image}, the local linear model (LLM) is proposed to exploit a DNN-based automatic compensation strategy for computing the quality score from local linear information. The deep similarity (DeepSim) \cite{gao2017deepsim} extracts feature maps for computing similarity from multiple layers of a pre-trained DNN for recognition, and then uses these recognition-related features to assess perceived quality. Bosse \emph{et al.} propose a DNN-based approach called deep image quality measure (DIQaM) \cite{bosse2018deep} to predict image visual quality.

\begin{figure}[t]
  \centerline{\includegraphics[width=7cm]{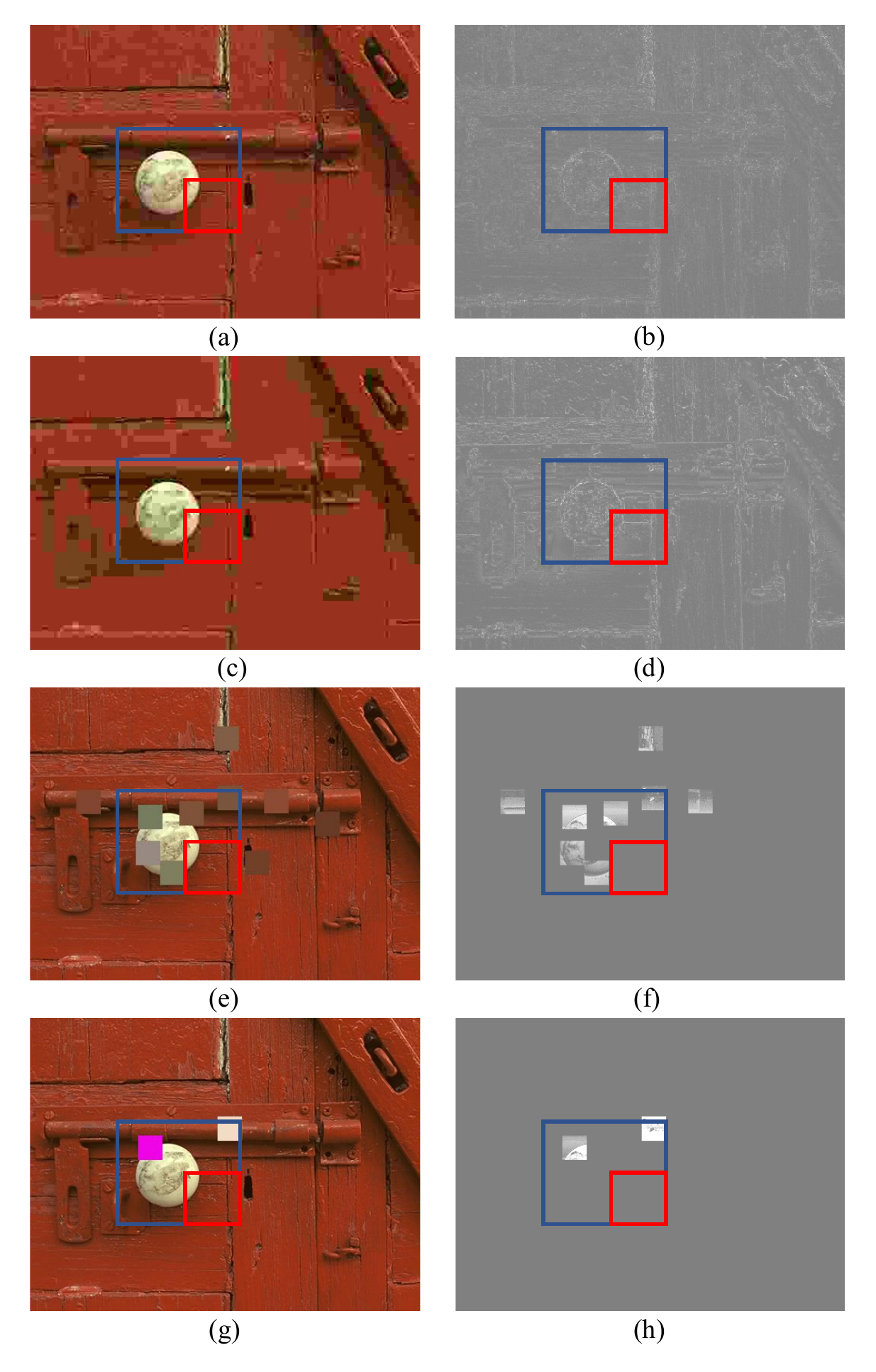}}
  \caption{(a) and (c) are examples of distorted images with different JPEG compression levels, while (e) and (g) are examples of distorted images with different sparse sampling and reconstruction levels. (b), (d), (f), and (h) are the corresponding residual maps between (a), (c), (e) as well as (g) and their reference image.}
  \centering
\label{fig1}
\end{figure}

Although existing IQA methods have demonstrated that DNNs are powerful visual models for learning discriminative features from image data, developing more efficient DNNs with the consideration of multi-scale feature representations for distorted image quality assessment is still prominent. As illustrated in Fig. \ref{fig1}, distorted images with various JPEG compression levels are shown in the left side of the first and second rows, which are selected from the TID2013 database \cite{ponomarenko2013color}. Their corresponding gray-scale residual maps computing from the reference image are shown in the right side. Another set of examples for distorted images with different sparse sampling and reconstruction levels are presented in the third and fourth rows. It can be seen that residual maps can well distinguish different distortion types and levels. Moreover, if we look at the local regions for the red bounding box of distorted images or residual maps, they are almost the same with each other under the same distortion type. But the local regions for the blue bounding box of distorted images or residual maps are quite different. Additionally, if we observe the whole images, they also vary widely. Therefore, we need to consider not only the effects of different scales at local regions, but also the influence of global features on perceptual visual quality. In other words, multi-scale features learning is important for image quality evaluation.

To overcome the limitation by only using one single scale of images, we resort to deep multi-scale features learning and incorporate such representations to an end-to-end DNN for better assessing the perceptual quality of distorted images. As described previously, except for distorted images, the residual maps between distorted and reference images are also discriminative to image visual quality. Moreover, inspired by the HVS is more sensitive to
luminance, both gray-scale residual maps and distorted images are exploited, which are input into two streams in the patch-wise manner. In our proposed network architecture, we apply two kinds of multi-scale feature representations, including spatial pyramid pooling and feature pyramid from the network structure. The spatial pyramid pooling considers both local and global spatial features, while the feature pyramid from the network structure focuses on the pyramidal hierarchy of DNNs at multi-scales. After that, fully connected layers are utilized to map multi-scale feature representations into the final scalar quality score. We train the whole network deeply supervised by perceptual quality scores. Extensive experimental results on four publicly available image quality databases validate that our proposed method outperforms state-of-the-arts. More importantly, we show the effectiveness of each component in the proposed network. Overall, our main contributions of this work are summarized as follows:

\begin{itemize}
  \item We propose a supervised DNN for the challenging distorted IQA problem. To our best knowledge, it is the first attempt to integrate multi-scale features learning into the end-to-end trained DNN architecture.
  \item Different from traditional multi-scale feature representations, we explore both spatial pyramid pooling and feature pyramid from the network structure. From these two multi-scale aspects, the proposed network is particularly suitable to evaluate the visual quality of distorted images with various distortion types and levels.
  \item Apart from distorted images, we use residual maps together with distorted images to construct two sub-networks. We combine pyramid learning with fully connected layers to map hierarchical multi-scale features to quality scores, achieving state-of-the-art performance on four image quality databases.
\end{itemize}

\section{Methodology}
\begin{figure*}[t]
  \centerline{\includegraphics[width=16cm]{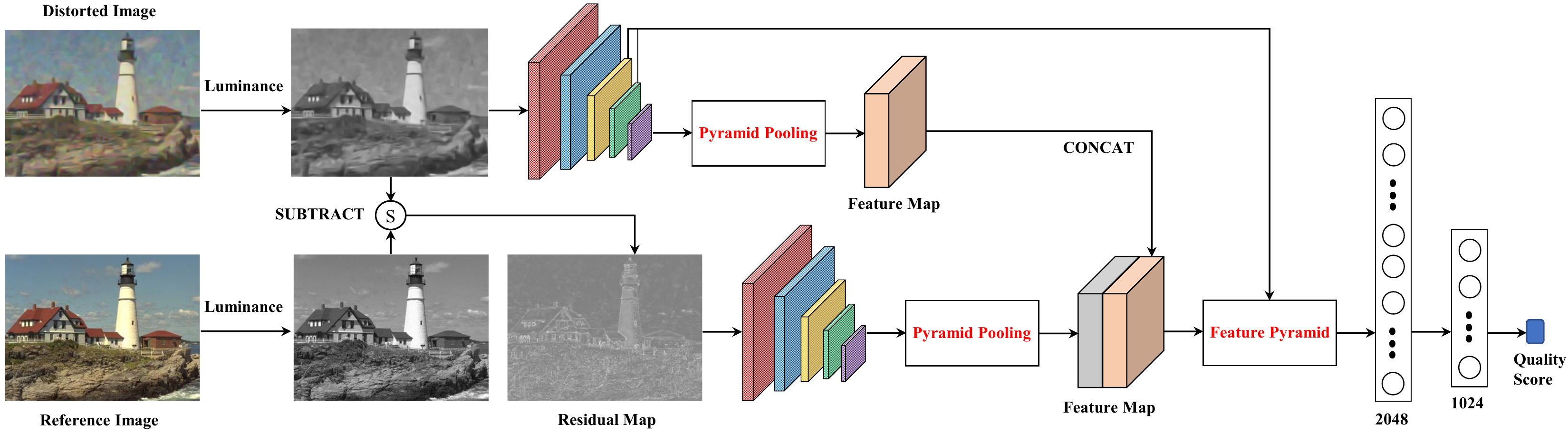}}
  \caption{Overall framework of our proposed model. It contains two streams respectively for distorted images and residual maps in luminance domain. The spatial pyramid pooling and the feature pyramid from the network structure are used to learn multi-scale feature representations.}
  \centering
\label{fig2}
\end{figure*}

We design an end-to-end optimized DNN for distorted image quality evaluation. Fig. \ref{fig2} depicts the overall framework of our proposed model. The proposed network contains two streams respectively for distorted images and the corresponding residual maps. Each sub-network takes an image patch of $128\times128$ as input and together predicts the overall perceived image visual quality as output. We first introduce spatial pyramid pooling to both two sub-networks. Then, the feature maps of two streams after pyramid pooling are concatenated to perform feature pyramid from the network structure. Finally, several fully connected layers are used to map the multi-scale features to perceptual image visual quality.

\subsection{Network Architecture}
As shown in Fig. \ref{fig2}, since the HVS is more sensitive to luminance, we convert the raw distorted and original reference images into the luminance domain, which are denoted by $\textbf{D}^{(n)}$ and $\textbf{O}^{(n)}$. $n = 1, ..., N$ represents the $n$-th image sample and $N$ is the total number of image samples. To obtain residual maps, a subtract operation is performed between the gray-scale distorted and original reference images:

\begin{equation}\label{1}
{\textbf{R}^{(n)}} = |{\textbf{D}^{(n)}} - {\textbf{O}^{(n)}}|.
\end{equation}

We denote training data as $\{({\textbf{D}^{(n)}}, {\textbf{R}^{(n)}}, {q^{(n)}})\} _{n = 1}^N$, where $\textbf{D}^{(n)}$ and $\textbf{R}^{(n)}$ are the $n$-th gray-scale distorted image and residual map, respectively. $q^{(n)}$ is the corresponding ground truth quality score of the inputs. Since our proposed method involves fully connected layers, the input image size should be fixed. Therefore, we use $128\times128$ image patches from gray-scale distorted images and residual maps to input into two sub-networks. Considering that the patch size is large to reflect the overall image visual quality, we set the ground truth quality score of each image patch as the corresponding image quality. In the whole network, we have two streams for distorted images and residual maps. Each sub-network is inspired by AlexNet \cite{krizhevsky2012imagenet}, which has 5 convolutional layers with max-pooling layers and the kernel size equalling to $3\times3$ \cite{simonyan2014very}. The kernel number is [8, 8, 16, 16, 32] for these five convolutional layers. Unlike the AlexNet, each convolutional layer is followed by the batch normalization (BN) \cite{ioffe2015batch} layer to alleviate gradient vanishing and data instability. Moreover, we exploit exponential linear unit (ELU) as the nonlinear activation function instead of conventional rectified linear unit (ReLU). Mathematically, the output of ELU is:

\begin{equation}\label{2}
elu(x)=\left\{
\begin{aligned}
& x, x\ge 0 \\
& \alpha({{e}^{x}}-1), x<0 \\
\end{aligned}
\right.,
\end{equation}
where $\alpha $ represents the parameter to control negative factors. In addition, the mean of the overall output is around zero, which is more robust than other activation functions.

\begin{figure}[t]
\centerline{\includegraphics[width=7cm]{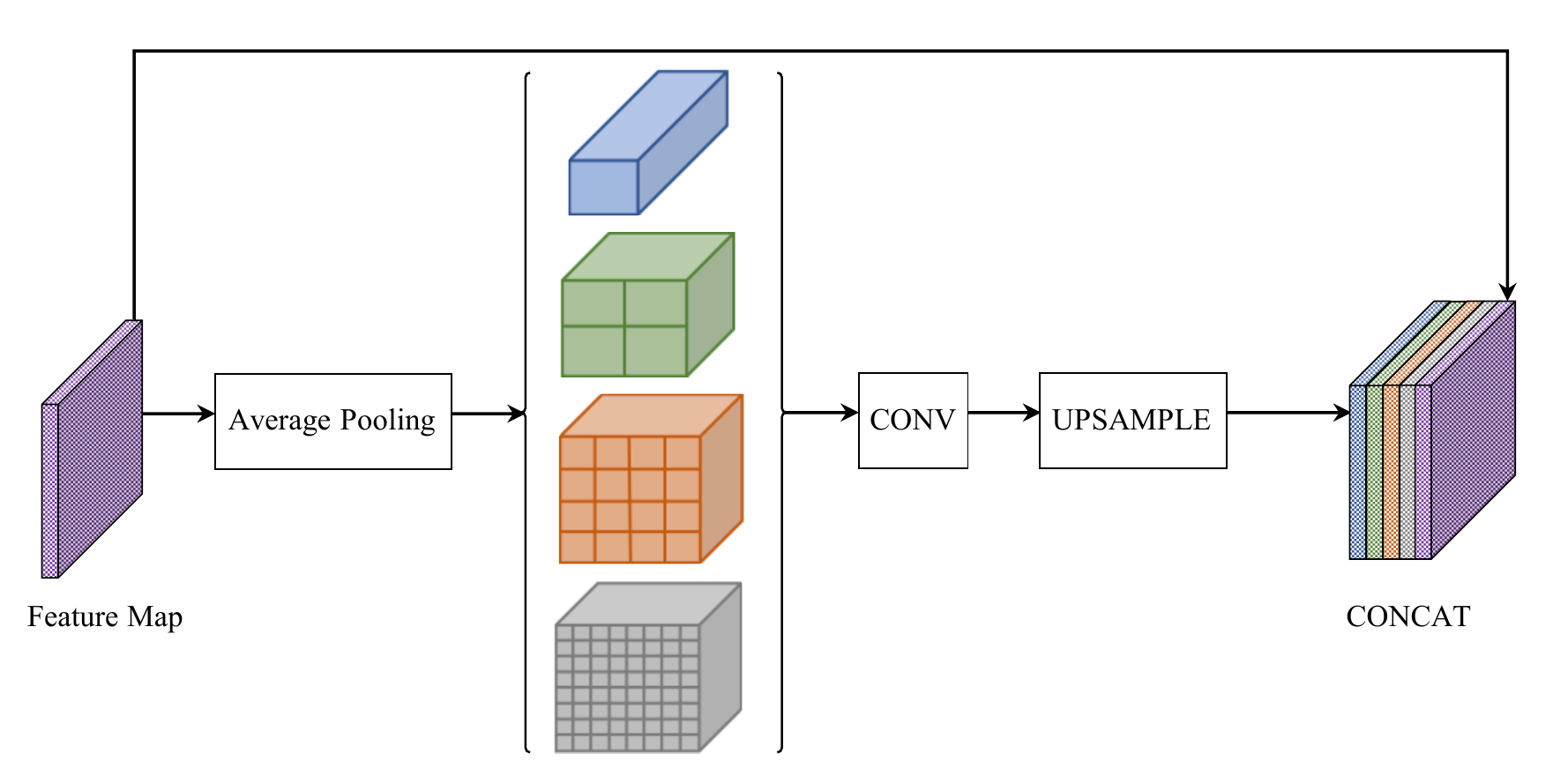}}
  \caption{Architecture of spatial pyramid pooling for each sub-network.}
\label{fig3}
\end{figure}

Afterward, we can obtain the output feature maps from the fifth convolutional layer of distorted image and residual map streams. Inspired by the HVS combines multi-scale features for perception, we aim to achieve multi-scale features learning in our proposed network. Specifically, the spatial pyramid pooling \cite{zhao2017pyramid} is used to the output feature maps of each sub-network as shown in Fig. \ref{fig3}. Our pyramid pooling is four-level with bin size as $1\times1$, $2\times2$, $4\times4$, and $8\times8$, which can capture both local and global spatial features with multi-scales. We concatenate the outputs of pyramid pooling modules, \emph{i.e.} $Concat(\mathcal{F}_D^n$, $\mathcal{F}_R^n$), and then apply feature pyramid \cite{lin2017feature} from the network structure of the distorted image stream. The feature pyramid aggregates multi-layers of the DNN structure aiming to exploit the inherent multi-scale, pyramidal hierarchy of DNNs. With the aid of spatial pyramid pooling for both two streams and feature pyramid from the network structure, multi-scale features are learned from the whole network. To evaluate the perceived image quality score, fully connected layers with size [2048, 1024, 1] are adopted after pyramid features learning.

\subsection{Training and Testing}
The proposed network is trained by back-propagation. The performance results are reported after 100 epochs. For the optimization of our network, the stochastic gradient descent (SGD) with momentum is adopted with an initial learning rate of ${{10}^{\text{-}3}}$. Moreover, the batch size, momentum factor and weight decay factor are fixed to 128, 0.9 and ${{10}^{\text{-}7}}$, respectively. We use $l_2$ loss to update the network weights as follows:

\begin{equation}\label{3}
{l_2}(\{ {\textbf{D}^{(n)}}, {\textbf{R}^{(n)}}\} ; \textbf{W}) = ||{q^{(n)}} - \mathop {{q^{(n)}}}\limits^ \wedge  |{|_2},
\end{equation}
where \textbf{W} denotes the network weights. $q^{(n)}$ and $\mathop {{q^{(n)}}}\limits^ \wedge$ are the ground truth subjective quality score and the network output, respectively.

During testing, given a testing distorted image, the residual map and distorted image in luminance domain are first computed. Then, we can predict its perceptual image visual quality by averaging the network outputs of each $128\times128$ image patch from the same image. Therefore, our proposed method effectively exploits the multi-scale feature representations of both distorted images and residual maps, further improve the end-to-end DNN for distorted image quality prediction.

\section{Experimental Results}
In this section, we show the results of comprehensive experiments on public databases to demonstrate the validity of our proposed quality assessment framework. In addition, we carry out a component analysis for better understanding of the proposed model.


\begin{table*}[t]
	\centering
    \scriptsize
	\caption{Performance Comparison Results on The LIVE, CSIQ, TID2008, and TID2013 Databases.}
    \scalebox{0.9}{
	\begin{tabular}{cc|ccccccc|cccc}
		\hline
        \hline
		Databases & Methods & PSNR & SSIM \cite{wang2004image} & MS-SSIM \cite{wang2003multiscale} & IW-SSIM \cite{wang2010information} & FSIMc \cite{zhang2011fsim} & VIF \cite{sheikh2006image} & SPSIM \cite{sun2018spsim} & LLM \cite{wang2016image} & DeepSim \cite{gao2017deepsim} & DIQaM-FR \cite{bosse2018deep} & Proposed \\ \hline
		\multirow{2}{*}{LIVE}
        & SRCC & 0.876 & 0.948 & 0.951 & 0.957 & 0.965 & 0.964 & 0.962 & 0.961 & 0.974 & 0.966 & \textbf{0.984} \\
		& PLCC & 0.872 & 0.945 & 0.949 & 0.952 & 0.961 & 0.960 & 0.960 & 0.958 & 0.968 & 0.977 & \textbf{0.984} \\ \hline
		\multirow{2}{*}{CSIQ}
        & SRCC & 0.801 & 0.876 & 0.913 & 0.921 & 0.931 & 0.919 & 0.944 & 0.905 & 0.919 & - & \textbf{0.960} \\
		& PLCC & 0.800 & 0.861 & 0.899 & 0.914 & 0.919 & 0.928 & 0.934 & 0.900 & 0.919 & - & \textbf{0.952} \\ \hline
		\multirow{2}{*}{TID2008}
        & SRCC & 0.525 & 0.775 & 0.854 & 0.860 & 0.884 & 0.749 & 0.910 & 0.908 & - & - & \textbf{0.928} \\
		& PLCC & 0.531 & 0.773 & 0.845 & 0.858 & 0.876 & 0.808 & 0.893 & 0.897 & - & - & \textbf{0.937} \\ \hline
		\multirow{2}{*}{TID2013}
        & SRCC & 0.639 & 0.742 & 0.786 & 0.778 & 0.851 & 0.677 & 0.904 & 0.904 & 0.846 & 0.859 & \textbf{0.906} \\
		& PLCC & 0.702 & 0.790 & 0.833 & 0.832 & 0.877 & 0.772 & 0.909 & 0.907 & 0.872 & 0.880 & \textbf{0.922} \\ \hline
        \hline
	\end{tabular}
}
\label{table2}
\end{table*}

\subsection{Benchmark Databases and Evaluation Criteria}
Four widely-used image quality benchmark databases are exploited for performance evaluation in our experiments, namely LIVE database \cite{sheikh2006statistical}, CSIQ database \cite{larson2010most}, TID2008 database \cite{ponomarenko2009tid2008}, and TID2013 database \cite{ponomarenko2013color}.

The LIVE database \cite{sheikh2006statistical} consists of 29 original reference images and 779 distorted images with 5 distortion types: white Gaussian noise (WN), Gaussian blur (BLUR), JPEG compression (JPEG), JPEG2000 compression (JP2K), and fast fading distortion (FF). Each distorted images is associated with a corresponding Differential Mean Opinion Score (DMOS) value in the range [0, 100]. The CSIQ database \cite{larson2010most} contains 30 original reference images and a total of 866 distorted images with 6 distortion types. This database is reported in the form of DMOS value normalized to the range [0, 1]. The TID2008 database \cite{ponomarenko2009tid2008} includes 25 reference images and 1700 distorted images with 17 distortion types at 4 degradation levels. The TID2013 database \cite{ponomarenko2013color} is an extension of TID2008, in which one more degradation level and 7 more distortions types are involved. In both TID2008 and TID2013 databases, MOS ranging in [0, 9] is provided for each distorted image.

We choose two commonly-applied criteria including the Spearman Rank Order Correlation Coefficient (SRCC) and Pearson Linear Correlation Coefficient (PLCC) to evaluate the performance of IQA approaches. SRCC is used to evaluate prediction monotonicity, while PLCC aims to evaluate prediction accuracy. For both correlation metrics, a higher correlation coefficient value indicates higher performance of a specific IQA method. Moreover, before calculating the PLCC performance, a four-parameter logistic function \cite{video2003final} is applied to map the predicted scores to the same scales of ground-truth quality scores.


\subsection{Performance Comparison}
\begin{figure}[t]
\centerline{\includegraphics[width=7cm]{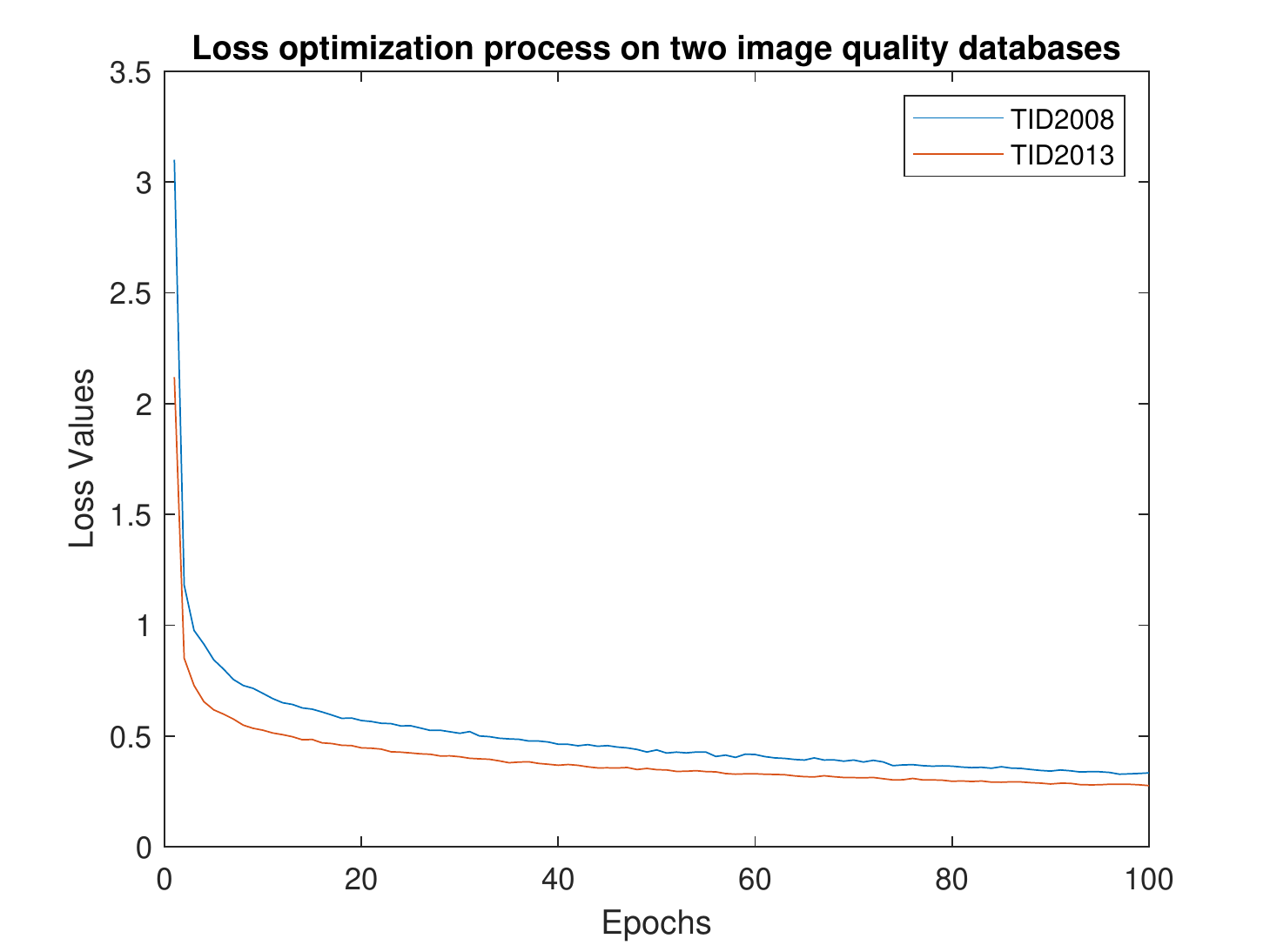}}
  \caption{Loss optimization process of the 100 epochs for the training set on the TID2008 and TID2013 databases.}
\label{fig4}
\end{figure}

We compare our proposed model with state-of-the-art approaches to validate its effectiveness. Each database is randomly divided into $80\%$ for training and $20\%$ for testing by the original reference images, and the corresponding distorted images are divided in the same way to ensure there is no content overlap between training and testing. The proposed method is compared with ten IQA models, \emph{i.e.}, PSNR, SSIM \cite{wang2004image}, MS-SSIM \cite{wang2003multiscale}, IW-SSIM \cite{wang2010information}, FSIMc \cite{zhang2011fsim}, VIF \cite{sheikh2006image}, SPSIM \cite{sun2018spsim}, LLM \cite{wang2016image}, DeepSim \cite{gao2017deepsim}, and DIQaM-FR \cite{bosse2018deep}. The performance results are listed in Table \ref{table2}, where the top performance in each row is highlighted in boldface. From this table, we can see that the proposed method outperforms state-of-the-art IQA models, especially for these deep learning-based algorithms, including the LLM \cite{wang2016image}, DeepSim \cite{gao2017deepsim} and DIQaM-FR \cite{bosse2018deep}. One possible explanation is that multi-scale characteristics are not taken into account in these DNN-based methods. In addition, since the quality score ranges of TID2008 and TID2013 are the same, the training loss optimization process of 100 epochs on these image quality databases is shown in Fig. \ref{fig4}. We can find that the training process of both two databases converges well.

\subsection{Ablation Study}

\begin{table}[t]
	\centering
    \scriptsize
	\caption{Ablation Study on The LIVE, CSIQ, TID2008, and TID2013 Databases.}
    \scalebox{0.9}{
	\begin{tabular}{cc|cccc}
		\hline
        \hline
		Databases & Methods &  Distorted Image & Residual Map & Direct Concat & Proposed \\ \hline
        \multirow{2}{*}{LIVE}
        & SRCC & 0.911 & 0.976 & 0.981 & \textbf{0.984} \\
        & PLCC & 0.913 & 0.978 & 0.981 & \textbf{0.984} \\ \hline
        \multirow{2}{*}{CSIQ}
        & SRCC & 0.745 & 0.917 & 0.949 & \textbf{0.960} \\
        & PLCC & 0.810 & 0.893 & 0.952 & \textbf{0.952} \\ \hline
        \multirow{2}{*}{TID2008}
        & SRCC & 0.529 & 0.894 & 0.923 & \textbf{0.928} \\
        & PLCC & 0.660 & 0.898 & 0.933 & \textbf{0.937} \\ \hline
        \multirow{2}{*}{TID2013}
        & SRCC & 0.577 & 0.854 & 0.900 & \textbf{0.906} \\
        & PLCC & 0.713 & 0.872 & 0.918 & \textbf{0.922} \\ \hline
        \hline
	\end{tabular}
}
\label{table3}
\end{table}

To demonstrate the effectiveness of our proposed multi-scale features learning, we conduct the ablation study. The performance results are shown in Table \ref{table3}, where the top performance in each row is highlighted in boldface. In this table, we compare the proposed method with several approaches, including a single stream with distorted images or residual maps as well as direct concatenation between the two streams. We can observe that using multi-scale features learning can effectively improve the DNN model. This is due to that the perceptual image visual quality is highly related to both global and local image regions. In addition, the inherent pyramidal hierarchy of the DNN model is also considered in our proposed method by feature pyramid.

\section{Conclusion}
In this paper, we tackle the challenging problem of distorted image quality prediction. We present an end-to-end optimized DNN with multi-scale features learning, which consists of two streams for distorted images and residual maps separately. The spatial pyramid pooling and feature pyramid from the network structure are considered in the architecture design, aiming to learn hierarchical multi-scale image feature representations. To obtain the ultimate perceived image quality, we combine the pyramid learning with fully connected layers for the quality score mapping. Experimental results show that our proposed method outperforms state-of-the-art IQA models on four commonly-used IQA databases and demonstrate the effectiveness of the proposed multi-scale components. In the future, we plan to extend deep multi-scale features learning to video quality assessment, where both spatial and temporal multi-scale features can be incorporated.

\bibliographystyle{IEEEbib}
\bibliography{references}

\end{document}